\documentclass{article}
\usepackage[utf8]{inputenc}
\usepackage[margin=0.95in]{geometry}
\usepackage{authblk}
\usepackage{amsmath}
\usepackage{amssymb}
\usepackage{hyperref}
\usepackage{float}
\usepackage{physics}
\usepackage{tikz}
\usepackage{slashed}
\usepackage{cite}
\usetikzlibrary{decorations.markings,decorations.pathmorphing}

\title{Manifestly Unitary Cosmological Perturbation Theory}

\author[1]{Panagiotis Christeas}
\author[1,2]{Logan Thomas}

\affil[1]{\textit{Department of Physics, Arizona State University, Tempe, AZ 85287}}
\affil[2]{\textit{Beyond: Center for Fundamental Concepts in Science,
Arizona State University, Tempe, AZ 85287}}

\date{}

\begin{document}

\maketitle

\abstract{The next decade will feature an abundance of novel cosmological data, while many fundamental questions about inflation remain.  Given this, there is ample need for maximally efficient calculations, especially in non-standard scenarios for the early Universe. In inflationary cosmology, observables are computed within the framework of in-in perturbation theory. Weinberg introduced a now-widely used re-organization of perturbation theory for in-in calculations. There is a subtle difference in the $i\epsilon$ prescriptions of Weinberg's perturbation series with traditional in-in, which could interfere with the projection onto the interacting vacuum. In this work, we show that a small adjustment to Weinberg's perturbation series yields agreement with standard in-in at every order of perturbation theory for commonly studied spins and masses in de Sitter spacetime. We then generalize the result to a large class of cosmological spacetimes, including slow roll spacetimes.}

\section{Introduction}

The quantum mechanics of fields in an expanding spacetime is of utmost importance in cosmology. Indeed, quantum fluctuations in the early universe plant the seeds for large-scale structures~\cite{Wang13}. The energy scales associated with inflation also make it a natural laboratory for high-energy physics~\cite{Nima15}. It is therefore desirable to have an efficient scheme to do quantum field theoretic computations on expanding backgrounds. Typically, the experimentally relevant observables are correlation functions, which are computed in the so-called in-in formalism.

In-in perturbation theory is primarily used in inflationary cosmology to compute correlation functions of operators, like the scalar and tensor fluctuations in the inflaton and metric, which are then imprinted onto the CMB. The most recent data show neither detection of non-Gaussianities~\cite{PlanckNG} nor the tensor power spectrum~\cite{PlanckPol}. As future experiments limit or detect these observables, constraints on physics during inflation tighten. Scalar~\cite{Nima15,Bartolo04,Seery07,Adshead09,Assassi12}, fermionic~\cite{Chung13,Kumar19}, vector~\cite{Bartolo12,Ferreira15,Lee16}, and spin-2~\cite{Ohashi13,Bordin18} fields all imprint their presence in the power spectra and non-Gaussianities in distinct ways, though they must be disentangled from the background contribution of the Standard Model~\cite{Chen10}, primordial black holes~\cite{Motohashi17,Passaglia18}, and the dynamics of inflation~\cite{Achucarro14,Cai18}. Alternative cosmologies for ultra-heavy dark matter could also modify the inflationary era with higher-spin states (see Ref.~\cite{SnowmassUHDM} and references therein).

The standard method to compute an in-in correlation function of some operator $\mathcal{O}$ is via
\begin{equation}
\langle\mathcal{O}(t)\rangle=\frac{\left\langle0\left|\left(\bar{T}e^{i\int_{-\infty_+}^{t}H_I\,dt'}\right)\mathcal{O}(t)\left(Te^{-i\int_{-\infty_-}^{t}H_I\,dt'}\right)\right|0\right\rangle}{\left\langle0\left|\left(\bar{T}e^{i\int_{-\infty_+}^{t}H_I\,dt'}\right)\left(Te^{-i\int_{-\infty_-}^{t}H_I\,dt'}\right)\right|0\right\rangle}\,,
\label{eq:inin}
\end{equation}
where $|0\rangle$ is the free-theory vacuum and $\infty_\pm=\infty(1\pm i\epsilon)$. Note that the $i\epsilon$'s, which project onto the interacting vacuum, spoil manifest unitarity. The forward and backward time-evolution operators no longer inverses of each other. The resulting diagrammatic formalism contains many redundancies related to the apparent violation of unitarity, which ultimately cancel out in physical quantities, as described in Ref.~\cite{Weinberg}. In that same work, the perturbation series was re-organized to remove the redundancies and put into a manifestly causal form involving nested commutators.
\begin{equation}
\langle\mathcal{O}(t)\rangle=\sum_{V=0}^{\infty}(-i)^V\int_{-\infty}^{t}\,dt_V\ldots\int_{-\infty}^{t_3}\,dt_2\int_{-\infty}^{t_2}\,dt_1\langle\Omega|[[\ldots[\mathcal{O}(t),H_I(t_V)],\ldots],H_I(t_1)]|\Omega\rangle\,,
\label{eq:weinberg}
\end{equation}
where $|\Omega\rangle$ is the interacting vacuum. Ignoring the $i\epsilon$'s of Eq.~\eqref{eq:inin}, one can derive Eq.~\eqref{eq:weinberg} by induction (See Appendix A of \cite{MattRaman19}). However, in the presence of the $i\epsilon$'s, the evolution of bra and ket are different, and therefore exchanging terms from the forward and backward time evolution operators, as in the commutators of Eq.~\eqref{eq:weinberg}, produces a different perturbation series. This was pointed out by Ref.s~\cite{Adshead09b,Senatore10} among others. For 15 years, this discrepancy remained unresolved. The first proof that these two perturbation series are equivalent at every order in perturbation theory for light scalars in exact de Sitter spacetime was given by~\cite{MattRaman}, which closed a loophole resent in a previous attempt~\cite{Kaya19}. However, a proof was only given for light scalars in exact de Sitter spacetime. In this paper, we will extend these works to show that these two perturbation series are equivalent order-by-order for commonly studied spins and all masses in a large class of FLRW spacetimes, including slow roll spacetimes.
In particular, we will show that the equivalence between the perturbation series \eqref{eq:inin} and a slightly modified Eq.~\eqref{eq:weinberg},
\begin{equation}
\langle\mathcal{O}(t)\rangle=\sum_{V=0}^{\infty}(-i)^V\int_{-\infty}^{t}\,dt_V\ldots\int_{-\infty}^{t_3}\,dt_2\int_{-\infty}^{t_2}\,dt_1\langle0|[[\ldots[\mathcal{O}(t),H_I^\epsilon(t_V)],\ldots],H_I^\epsilon(t_1)]|0\rangle\,,
\label{eq:manifestUnitaryPT}
\end{equation}
where $H_I^\epsilon(t)\equiv H_I(t)e^{\epsilon t}$, and we have replaced the interacting vacuum with the free vacuum.

The remainder of this paper is organized as follows. In the following section, we will show how these series agree at every order for scalars (Sec.~\ref{subsec:scalar}), spin-1/2 fermions (Sec.~\ref{subsec:fermion}), vectors (Sec.~\ref{subsec:vector}), and gravitons (Sec.~\ref{subsec:graviton}) explicitly in de Sitter spacetime. We end in Sec.~\ref{sec:conclusion} with a short discussion of our results and the generalization to past-infinite FLRW spacetimes.

\section{Spin-0 and general argument}\label{subsec:scalar}

In this section, we will use the scalar case, reproduced from Ref.~\cite{MattRaman}, as an example to show the general argument, which we will later adapt for other spins.\footnote{In Ref.~\cite{MattRaman}, the authors assume minimally coupled scalars with $m < 3H/2$. As we shall see, the same argument holds for all masses and non-minimal couplings.} We will consider the general case of non-minimally coupled massive scalars, described by the action
\begin{equation}
\mathcal{S}_0=\int\,d^dx\sqrt{-g}\frac{1}{2}\left[\nabla_\mu\phi\nabla^\mu\phi-m^2\phi^2-\xi R\phi^2\right]\,.
\end{equation}
This leads to an equation of motion (in conformal coordinates)
\begin{equation}
\partial_0{}^2\phi+2aH\partial_0\phi+(k^2+a^2m^2+\xi a^2R)\phi=0\,.
\label{eq:scalarEoM}
\end{equation}
The solution obeying the Bunch-Davies boundary conditions is
\begin{equation}
\phi_{\vec{k}}(\tau) = \frac{H}{\sqrt{2}(2\pi)^{3/2}}\,\tau^{3/2}H^{(2)}_\nu(-k\tau)\,,
\end{equation}
where $\nu^2=(d-1)^2/4-m^2/H^2-d(d-1)\xi$. For $\nu$ a half odd integer, e.g. in the massless minimally ($\xi=0$) or conformally ($\xi=1/6$) coupled cases, the Hankel function reduces to an elementary function. In general, these Hankel functions are difficult to deal with, however we only need the early-time behavior for our purposes. The asymptotic expansions of the Hankel functions $H_\nu^{(1,2)}(z)$ as $z\rightarrow\infty$ are
\begin{align}
H^{(1)}_\nu(z) &\sim \left(\frac{2}{\pi z}\right)^{1/2}e^{i(z-\frac{1}{2}\pi\nu-\frac{1}{4}\pi)}\sum_{n=1}^{\infty}i^n\frac{a_n(\nu)}{z^n}\nonumber\\
H^{(2)}_\nu(z) &\sim \left(\frac{2}{\pi z}\right)^{1/2}e^{-i(z-\frac{1}{2}\pi\nu-\frac{1}{4}\pi)}\sum_{n=1}^{\infty}(-i)^n\frac{a_n(\nu)}{z^n}\,,
\label{eq:asymptoticHankel}
\end{align}
where the $a_n(\nu)$ are numerical coefficients that can be found in any standard reference (e.g. Sect. 10 of~\cite{DLMF}). The above asymptotic expansions are valid for any complex $\nu$, which allows this argument to hold for large masses and non-minimal couplings, for which $\nu^2<0$.

Within in-in perturbation theory, one encounters 4 different propagators, depending on whether the field operators originate from the forward ($+$) or backward ($-$) time-evolution operators.
\begin{align}
G^{++}(\tau,\tau') &= \langle T\phi^{+}(\tau)\phi^{+}(\tau')\rangle\nonumber\\
G^{--}(\tau,\tau') &= \langle \bar{T}\phi^{-}(\tau)\phi^{-}(\tau')\rangle\nonumber\\
G^{-+}(\tau,\tau') &= \langle \phi^{-}(\tau)\phi^{+}(\tau')\rangle\nonumber\\
G^{+-}(\tau,\tau') &= \langle \phi^{-}(\tau')\phi^{+}(\tau)\rangle,
\label{eq:propagators}
\end{align}
where $-$ operators are always to the left of $+$ operators, as required by the form of Eq.~\eqref{eq:inin}, and we evaluate these all in the free vacuum. The $i\epsilon$ prescription can be incorporated by deforming the time argument of the field operators to $\tau(1-i\epsilon)$ for $\phi^+$ and $\tau(1+i\epsilon)$ for $\phi^-$.
\begin{align}
\phi^+_{\vec k} &= \frac{H}{\sqrt{2}(2\pi)^{3/2}}\left[\tau^{3/2}H^{(2)}_\nu(k\tau(1-i\epsilon))a_{\vec k}+\tau^{3/2}H_\nu^{(1)}(k\tau(1-i\epsilon))a_{\vec k}^\dagger\right]\nonumber\\
\phi^-_{\vec k} &= \frac{H}{\sqrt{2}(2\pi)^{3/2}}\left[\tau^{3/2}H^{(2)}_\nu(k\tau(1+i\epsilon))a_{\vec k}+\tau^{3/2}H_\nu^{(1)}(k\tau(1+i\epsilon))a_{\vec k}^\dagger\right]
\end{align}
Because the $i\epsilon$ prescription only matters at early times, it is sufficient to include only the dominant contribution from the asymptotic expansions in Eq.~\eqref{eq:asymptoticHankel}, that is the exponential contribution. We can then write a new set of $\phi^\pm$ fields which will yield the same results as the above in the limits of early time and small $\epsilon$.
\begin{align}
\phi^+_{\vec k} &\rightarrow \frac{H}{\sqrt{2}(2\pi)^{3/2}}\left[e^{-\epsilon\tau}\tau^{3/2}H^{(2)}_\nu(k\tau)a_{\vec k}+e^{\epsilon\tau}\tau^{3/2}H_\nu^{(1)}(k\tau)a_{\vec k}^\dagger\right]\nonumber\\
\phi^-_{\vec k} &\rightarrow \frac{H}{\sqrt{2}(2\pi)^{3/2}}\left[e^{\epsilon\tau}\tau^{3/2}H^{(2)}_\nu(k\tau)a_{\vec k}+e^{-\epsilon\tau}\tau^{3/2}H_\nu^{(1)}(k\tau)a_{\vec k}^\dagger\right]
\end{align}
From these fields, we can build the various in-in propagators of Eq.~\eqref{eq:propagators}
\begin{align}
G^{++}(\tau,\tau') &= \frac{H^2(\tau\tau')^{3/2}}{2}\left[\Theta(\tau-\tau')e^{-\epsilon(\tau-\tau')}H^{(1)}_\nu(k\tau')H^{(2)}_\nu(k\tau) + \Theta(\tau'-\tau)e^{-\epsilon(\tau'-\tau)}H^{(1)}_\nu(k\tau)H^{(2)}_\nu(k\tau')\right]\nonumber\\
G^{--}(\tau,\tau') &= \frac{H^2(\tau\tau')^{3/2}}{2}\left[\Theta(\tau-\tau')e^{-\epsilon(\tau-\tau')}H^{(1)}_\nu(k\tau)H^{(2)}_\nu(k\tau') + \Theta(\tau'-\tau)e^{-\epsilon(\tau'-\tau)}H^{(1)}_\nu(k\tau')H^{(2)}_\nu(k\tau)\right]\nonumber\\
G^{-+}(\tau,\tau') &= \frac{H^2(\tau\tau')^{3/2}}{2}e^{\epsilon(\tau+\tau')}H^{(1)}_\nu(k\tau')H^{(2)}_\nu(k\tau)\nonumber\\
G^{+-}(\tau,\tau') &= \frac{H^2(\tau\tau')^{3/2}}{2}e^{\epsilon(\tau+\tau')}H^{(1)}_\nu(k\tau)H^{(2)}_\nu(k\tau')\,.
\label{eq:explicitScalarProps}
\end{align}
In what follows, the exponential time dependence of each of these propagators will be of particular importance
\begin{align}
G^{++}(\tau,\tau') &\propto e^{-\epsilon|\tau-\tau'|}\nonumber\\
G^{--}(\tau,\tau') &\propto e^{-\epsilon|\tau-\tau'|}\nonumber\\
G^{-+}(\tau,\tau') &\propto e^{\epsilon(\tau+\tau')}\nonumber\\
G^{+-}(\tau,\tau') &\propto e^{\epsilon(\tau+\tau')}\,.
\label{eq:scalarPropBehavior}
\end{align}
We can likewise construct the manifestly unitary perturbation series of Eq.~\eqref{eq:manifestUnitaryPT} by deforming our field by a factor $e^{\epsilon\tau}$, sending $\phi^+$ and $\phi^-$ to $e^{\epsilon\tau}\phi$. This leads to propagators similar to those in Eq.~\eqref{eq:explicitScalarProps}, only differing in their exponential dependence on $\tau$ and $\tau'$,
\begin{equation}
G_U^{\pm\pm} \propto e^{\epsilon(\tau+\tau')}\,.
\end{equation}
With our propagators in Eq.~\eqref{eq:explicitScalarProps}, we can consider a generic diagram, decomposed into parts where the vertices are time-ordered
\begin{multline}
\langle\mathcal{O}(\tau)\rangle \supset \lim_{\epsilon\rightarrow0}\lim_{T\rightarrow-\infty}\int\frac{d^3k_1}{(2\pi)^3}\ldots f(k_1,\ldots)\\\times\int_T^\tau\,d\tau_V F_V(\tau_V)\int_{\tau_V}^\tau\,d\tau_{V-1}F_{V-1}(\tau_{V-1})\ldots\int_{\tau_2}^\tau\,d\tau_1 F_1(\tau_1)F_0(\tau)\,.
\label{eq:genericTerm}
\end{multline}
Here, $\supset$ means that the correlator contains this term in its perturbation series (as well as other time orderings of the vertices), $f(k_1,\ldots)$ contains all of the constants and $k_i$-dependence that can be factored out of the time integrals, and each of the $F_j(\tau_j)$ is a product of Hankel functions and factors of $\tau_j$ coming from factors of the metric and measure. The form of $F_j(\tau_j)$ then implies the asymptotic expansion
\begin{equation}
F_j(\tau_j)\sim\sum_n b^{(j)}_n\frac{e^{i\omega_j^\epsilon \tau_j}}{\tau_j{}^n}\,,
\label{eq:FjAsymptotic}
\end{equation}
where $\omega_j^\epsilon$ is a linear combination of $k$ and $i\epsilon$ determined by the forms of Eqs.~\eqref{eq:asymptoticHankel} and \eqref{eq:scalarPropBehavior}. Functions having the asymptotic form in Eq.~\eqref{eq:FjAsymptotic} have integrals with asymptotic form
\begin{equation}
\int_{\tau_i}^\tau\,d\tau_j F(\tau_j) \sim c_j(\tau) + \sum_n d_n\frac{e^{i\omega^\epsilon_j\tau_i}}{\tau_i{}^n}\,,
\end{equation}
for $\tau_i\rightarrow-\infty$, where $c_j(\tau)$ can be thought of as an integration constant. Therefore, each integral in Eq.~\eqref{eq:genericTerm} will return a function with the same asymptotic form. This allows us to resolve the integrals.
\begin{equation}
\langle\mathcal{O}(\tau)\rangle \supset \lim_{\epsilon\rightarrow0}\lim_{T\rightarrow-\infty}\int\frac{d^3k_1}{(2\pi)^3}\ldots f(k_1,\ldots)\sum_{i=0}^{V}R_i(\omega_1,\ldots;\tau)\sum_{N=0}^{\infty}\frac{C_{N}^{i}e^{\sum_{n=1}^{i}i\omega_n \tau}e^{\sum_{m=i+1}^{V}i\omega_m^\epsilon T}}{T^{N}}
\label{eq:genericDiagram}
\end{equation}
Notice that we have neglected any $\epsilon$ factors except those multiplying $T$ in the exponential. Since we are taking the $T\rightarrow-\infty$ limit, we only need to consider the $\epsilon$ suppression from this factor. Each term in the sum in Eq.~\eqref{eq:genericDiagram} accounts for a different choice for which integration limit to follow at each integral, having only $V+1$ terms due to the time-ordering of the vertices. Due to the mixed behavior of the propagators (Eq.~\eqref{eq:scalarPropBehavior}), one might wonder if there is the possibility of exponential enhancement in the $T\rightarrow-\infty$ limit of Eq.~\eqref{eq:genericDiagram}. To show that this is not the case, we can observe that as a result of the integrals, every vertex time is evaluated either at $\tau$ or $T$. Therefore, we should examine the early-$T$ behavior of the propagators with these endpoints,
\begin{align}
\left\{G^{\pm\pm}(\tau,T),G^{\pm\pm}(T,\tau)\right\} &\propto e^{-\epsilon|\tau-T|} \sim e^{\epsilon T}\nonumber\\
\left\{G^{\pm\pm}(\tau,\tau),G^{\pm\pm}(T,T)\right\} &\propto e^{0} \sim 1\nonumber\\
\left\{G^{\pm\mp}(\tau,T),G^{\pm\mp}(T,\tau)\right\} &\propto e^{\epsilon(\tau+T)} \sim e^{\epsilon T}\nonumber\\
G^{\pm\mp}(\tau,\tau) &\propto e^{\epsilon(2\tau)} \sim 1\nonumber\\
G^{\pm\mp}(T,T) &\propto e^{\epsilon(2T)} \sim e^{2\epsilon T}\,.
\end{align}
The worst we can do is to have no enhancement at all. But note further that for terms in Eq.~\eqref{eq:genericDiagram} with $i<V$, any diagram that connects to the correlation time must have a contribution from propagator that goes between $\tau$ and $T$, which always gives an exponential suppression. A diagram which does not connect to the correlation time is a bubble diagram, which does not contribute anyway, since we divide out bubble diagrams in Eq.~\eqref{eq:inin}. We can then take the $T\rightarrow-\infty$ limit of Eq.~\eqref{eq:genericDiagram} to find the much simpler result
\begin{equation}
\langle\mathcal{O}(\tau)\rangle \supset \int\frac{d^3k_1}{(2\pi)^3}\ldots f(k_1,\ldots)R_V(\omega_1,\ldots;\tau)C_{0}^{V}e^{\sum_{n=1}^{V}i\omega_n \tau}.
\label{eq:genericDiagramLimit}
\end{equation}
Let us now compare Eq.~\eqref{eq:genericDiagramLimit} to the same computation done with the propagators $G_U$. In this case, where we have uniform $\epsilon$ dependence, the $T\rightarrow-\infty$ limit in Eq.~\eqref{eq:genericDiagram} can immediately be computed to obtain Eq.~\eqref{eq:genericDiagramLimit}. The $\epsilon$ suppression from every $G_U$ propagator also implies that any diagram that is not connected to the correlation time will only depend on $T$, and thus will be removed by the $T\rightarrow-\infty$ limit. Therefore, the perturbation series generated by Eqs. \eqref{eq:inin} and \eqref{eq:manifestUnitaryPT} are equivalent.\\
\indent Before continuing on to other spins, it is worth reflecting on what features of this proof are specific to the scalar nature of the field. Of utmost importance was the asymptotic expansion of the mode functions, however the important features of those expansions are enforced by the Bunch-Davies boundary conditions. While the equation of motion, and thus the mode functions, will differ by the spin of the particle, the Bunch-Davies boundary conditions ensure that the asymptotic expansion of those mode functions all have the same general form. This hints at the generality of this argument, but we still need to account for any spin-related structures that might appear in correlators of fermions, vectors, or gravitons.

\section{Higher spin fields}\label{sec:higherSpin}

\subsection{Spin-1/2}\label{subsec:fermion}

The quadratic action for the massive Dirac fermion is
\begin{equation}
\mathcal{S}_{1/2} = \int\,d^dx\,e\left[\frac{i}{2}\left(\bar{\Psi}\Gamma^\mu\nabla_\mu\Psi-\nabla_\mu\bar{\Psi}\Gamma^\mu\Psi\right)-m\bar{\Psi}\Psi\right]\,,
\end{equation}
where $e_\alpha{}^\mu$ is a veilbein, $\Gamma^\mu=e_\alpha{}^\mu\gamma^\alpha$, and $e$ is the determinant of $e_\alpha{}^\mu$. This gives an equation of motion\cite{Fermionloops}
\begin{equation}
(i\Gamma^\mu\nabla_\mu-m)\Psi=0\,.
\end{equation}
The massless fermion is conformally invariant, with conformal weight $3/2$, which motivates the redefinition $\psi\equiv a^{3/2}(\tau)\Psi$. Written explicitly in conformal FLRW coordinates with a scale factor $a$, the equation of motion for $\psi$ is
\begin{equation}
i\gamma^\mu\partial_\mu\psi-am\psi=0\,.
\end{equation}
%[Write something useful here]
The spinors and anti-spinors can be written in terms of creation and annihilation operators as:

\begin{align}
	\psi(x) = \int\dfrac{d^3k}{(2\pi)^3}\left[ \hat{a}_{\vec k}\chi(k)e^{i\vec{k}\cdot\vec{x}} + \hat{b}^\dagger_{\vec k}\nu(k)e^{-i\vec{k}\cdot\vec{x}}\right]\, \nonumber\\
	\bar{\psi}(x) = \int\dfrac{d^3k}{(2\pi)^3} \left[\hat{b}_{\vec k}\bar{\nu}(k)e^{i\vec{k}\cdot\vec{x}} + \hat{a}^\dagger_{\vec k}\bar{\chi}(k)e^{-i\vec{k}\cdot\vec{x}}\right]
\end{align}

\noindent We will briefly summarize the quantization of a fermion in FLRW spacetime, as it is more involved than the scalar case. For further details, we refer the reader to Ref.~\cite{FermionProp}. Writing the spinor and anti-spinor in helicity basis
\begin{align}
	&\chi(k,\tau)=\sum_h \left( \begin{array}{c}
	\chi_{L,h}(k,\tau) \\
	\chi_{R,h}(k,\tau)
	\end{array}\right)\otimes \xi_h\nonumber\\
	&\nu(k,\tau)=\sum_h \left( \begin{array}{c}
	\nu_{R,h}(k,\tau) \\
	\nu_{L,h}(k,\tau)
	\end{array}\right)\otimes \xi_h\,,
\end{align}
where $L$ and $R$ stands for left- and right-handed respectively, and $\xi_h$ is a 2-component helicity eigenspinor, satisfying $\vec{k}\cdot\vec{\sigma}\xi_h=hk\xi_{h}$. The decomposition of the spinor in the helicity basis turns the Dirac equation into the following coupled system of differential equations:
\begin{align}
	&i \dfrac{d}{d\tau} \chi_{L,h} + hk \chi_{L,h} - am\chi_{R,h} = 0\nonumber\\
	&i \dfrac{d}{d\tau} \chi_{R,h} - hk \chi_{R,h} - am\chi_{L,h} = 0\,.
\label{eq:chiDiffEq}
\end{align}
The same equations apply for $\nu_{L,h}$ and $\nu_{R,h}$. The Bunch-Davies boundary conditions for $\chi_{L,R}$ and $\nu_{L,R}$ are
\begin{align}
	\chi_{L,h} &\sim \frac{1-h}{2}e^{-ik\tau}\nonumber\\
	\chi_{R,h} &\sim \frac{1+h}{2}e^{-ik\tau}\nonumber\\
	\nu_{L,h} &\sim \frac{1+h}{2}e^{ik\tau}\nonumber\\
	\nu_{R,h} &\sim \frac{1-h}{2}e^{ik\tau}\,.
\label{eq:fermionBC}
\end{align}
The mode functions satisfying both Eqs~\eqref{eq:chiDiffEq} and \eqref{eq:fermionBC} in de Sitter are
\begin{align}
 &\chi_{L,h}(k,\tau)=\sqrt{-\dfrac{\pi k\tau}{8}} \left[e^{i\dfrac{\pi}{2}\left(\nu_+ + 1/2\right)}H^{(1)}_{\nu_+}(-k\tau)-he^{i\dfrac{\pi}{2}\left(\nu_- + 1/2\right)}H^{(1)}_{\nu_-}(-k\tau)\right]\nonumber\\
  &\chi_{R,h}(k,\tau)=\sqrt{-\dfrac{\pi k\tau}{8}} \left[e^{i\dfrac{\pi}{2}\left(\nu_+ + 1/2\right)}H^{(1)}_{\nu_+}(-k\tau)+he^{i\dfrac{\pi}{2}\left(\nu_- + 1/2\right)}H^{(1)}_{\nu_-}(-k\tau)\right]\nonumber\\
   &\nu_{L,h}(k,\tau)=\sqrt{-\dfrac{\pi k\tau}{8}} \left[e^{-i\dfrac{\pi}{2}\left(\nu_+ + 1/2\right)}H^{(2)}_{\nu_+}(-k\tau)+he^{-i\dfrac{\pi}{2}\left(\nu_- + 1/2\right)}H^{(2)}_{\nu_-}(-k\tau)\right]\nonumber\\
    &\nu_{R,h}(k,\tau)=\sqrt{-\dfrac{\pi k\tau}{8}} \left[e^{-i\dfrac{\pi}{2}\left(\nu_+ + 1/2\right)}H^{(2)}_{\nu_+}(-k\tau)-he^{-i\dfrac{\pi}{2}\left(\nu_- + 1/2\right)}H^{(2)}_{\nu_-}(-k\tau)\right]\,,
\end{align}
where $\nu_{\pm} \equiv \dfrac{1}{2} \mp i\dfrac{m}{H}$. Just as in the scalar case, we use the $i\epsilon$ prescription and keep the dominant exponential contribution. Thus, at early time and small $\epsilon$ the $\chi$'s and $\nu$'s will be:
\begin{align}
	 &\chi_{L,h}\left(k\tau(1\pm i\epsilon)\right)\rightarrow\sqrt{-\dfrac{\pi k\tau}{8}} e^{\mp \epsilon \tau}\left[e^{i\dfrac{\pi}{2}\left(\nu_+ + 1/2\right)}H^{(1)}_{\nu_+}(-k\tau)-he^{i\dfrac{\pi}{2}\left(\nu_- + 1/2\right)}H^{(1)}_{\nu_-}(-k\tau)\right]\nonumber\\
  &\chi_{R,h}\left(k\tau(1\pm i\epsilon)\right)\rightarrow\sqrt{-\dfrac{\pi k\tau}{8}} e^{\mp \epsilon \tau}\left[e^{i\dfrac{\pi}{2}\left(\nu_+ + 1/2\right)}H^{(1)}_{\nu_+}(-k\tau)+he^{i\dfrac{\pi}{2}\left(\nu_- + 1/2\right)}H^{(1)}_{\nu_-}(-k\tau)\right]\nonumber\\
   &\nu_{L,h}\left(k\tau(1\pm i\epsilon)\right)\rightarrow\sqrt{-\dfrac{\pi k\tau}{8}} e^{\pm \epsilon \tau}\left[e^{-i\dfrac{\pi}{2}\left(\nu_+ + 1/2\right)}H^{(2)}_{\nu_+}(-k\tau)+he^{-i\dfrac{\pi}{2}\left(\nu_- + 1/2\right)}H^{(2)}_{\nu_-}(-k\tau)\right]\nonumber\\
    &\nu_{R,h}\left(k\tau(1\pm i\epsilon)\right)\rightarrow\sqrt{-\dfrac{\pi k\tau}{8}} e^{\pm \epsilon \tau}\left[e^{-i\dfrac{\pi}{2}\left(\nu_+ + 1/2\right)}H^{(2)}_{\nu_+}(-k\tau)-he^{-i\dfrac{\pi}{2}\left(\nu_- + 1/2\right)}H^{(2)}_{\nu_-}(-k\tau)\right]
\end{align}
This will allow us to write the four in-in propagators for a fermion. We define the following in anticipation of the results for the propagators
\begin{align}
\chi_a(\tau)\bar{\chi}_b(\tau')&=\hat{\Delta}_{ab}\left[\left(\frac{1+\gamma^0}{2}\right)\sqrt{\tau\tau'} H_{\nu_-}^{(1)}(-k\tau) H_{\nu_-}^{(2)}(-k\tau') + \left(\frac{1-\gamma^0}{2}\right)\sqrt{\tau\tau'} H_{\nu_+}^{(1)}(-k\tau) H_{\nu_+}^{(2)}(-k\tau')\right]\nonumber\\
\nu_a(\tau)\bar{\nu}_b(\tau')&=-\hat{\Delta}_{ab}\left[\left(\frac{1+\gamma^0}{2}\right)\sqrt{\tau\tau'} H_{\nu_-}^{(2)}(-k\tau) H_{\nu_-}^{(1)}(-k\tau') + \left(\frac{1-\gamma^0}{2}\right)\sqrt{\tau\tau'} H_{\nu_+}^{(2)}(-k\tau) H_{\nu_+}^{(1)}(-k\tau')\right]
\label{eq:chichibar}
\end{align}
where we have defined the operator $\hat{\Delta}\equiv\frac{\pi}{4}a^{-3/2}(\tau)a^{-3/2}(\tau')[i\gamma^0_{ab}\partial_0-\gamma^i_{ab}k_i + a(\tau)m]$ and explicitly written the spinor indices. These definitions allow us to write the propagators as
\begin{align}
G^{++}_{ab}(\tau,\tau') &= \langle T\{\psi^+_a(\tau)\bar{\psi}^+_b(\tau')\}\rangle =\Theta(\tau-\tau')e^{-\epsilon(\tau-\tau')}\chi_a(\tau)\bar{\chi}_b(\tau') -\Theta(\tau'-\tau)e^{-\epsilon(\tau'-\tau)}\nu_a(\tau)\bar{\nu}_b(\tau')\nonumber\\
G^{--}_{ab}(\tau,\tau')&= \langle \bar{T}\{\psi^-_a(\tau)\bar{\psi}^-_b(\tau')\}\rangle =\Theta(\tau'-\tau)e^{-\epsilon(\tau'-\tau)}\chi_a(\tau)\bar{\chi}_b(\tau') -\Theta(\tau-\tau')e^{-\epsilon(\tau-\tau')}\nu_a(\tau)\bar{\nu}_b(\tau')\nonumber\\
G^{+-}_{ab}(\tau,\tau') &= \langle\bar{\psi}^-_b(\tau')\psi^+_a(\tau)\rangle = e^{\epsilon(\tau+\tau')}\nu_a(\tau)\bar{\nu}_b(\tau')\nonumber\\
G^{-+}_{ab}(\tau,\tau') &= \langle\psi^-_a(\tau)\bar{\psi}^+_b(\tau')\rangle = e^{\epsilon(\tau+\tau')}\chi_a(\tau)\bar{\chi}_b(\tau')
\end{align}
The asymptotic behavior of these propagators, with respect to $\epsilon$, is the same as in the scalar case, Eq.~\eqref{eq:scalarPropBehavior}
\begin{align}
G^{++}_{ab}(\tau,\tau') &\propto e^{-\epsilon|\tau-\tau'|}\nonumber\\
G^{--}_{ab}(\tau,\tau') &\propto e^{-\epsilon|\tau-\tau'|}\nonumber\\
G^{-+}_{ab}(\tau,\tau') &\propto e^{\epsilon(\tau+\tau')}\nonumber\\
G^{+-}_{ab}(\tau,\tau') &\propto e^{\epsilon(\tau+\tau')}\,.
\end{align}
From this point, the same arguments made in the scalar case hold in the fermion case. The non-trivial spinor structure in Eq.~\eqref{eq:chichibar} is time-independent, except for the scale factor appearing in the mass term. This clearly goes to 0 at early times, and so can be ignored. The propagators are products of Hankel functions and derivatives of Hankel functions, which all obey the same general properties we need at early times in order to ensure that the two perturbation series agree at all orders.

\subsection{Spin-1}\label{subsec:vector}

The quadratic spin-1 action is
\begin{equation}
\mathcal{S}_{1}=\int\,d^dx\sqrt{-g}\left[-\frac{1}{4}F_{\mu\nu}^2-\frac{1}{2}m^2A_\mu A^\mu\right]\,.
\end{equation}
It is useful to separate out the spatial and temporal components of $A_\mu$ so that we can track the physical degrees of freedom. After some work, we can rewrite the action as~\cite{KolbLong20}
\begin{multline}
\mathcal{S}_1=\int\,d\tau\int\frac{d^3k}{(2\pi)^3}\frac{1}{2}\bigg[(k^2+a^2m^2)\left|A_0+i\frac{k_i\partial_0A_i}{k^2+a^2m^2}\right|^2+|\partial_0A_i|^2-\frac{|k_i\partial_0A_i|^2}{k^2+a^2m^2}\\+\frac{1}{2}|k_iA_j-k_jA_i|^2-a^2m^2|A_i|^2\bigg]\,.
\label{eq:explicitVectorAction}
\end{multline}
The temporal component of the field does not have a kinetic term, and is therefore non-dynamical and can be integrated out. It is clear that to minimize the action with respect to $A_0$, we set
\begin{equation}
A_0=-i\frac{k_i\partial_0A_i}{k^2+a^2m^2}\,.
\label{eq:A0Solution}
\end{equation}
We can decompose the field into its three polarizations (in the massive case), which have action
\begin{align}
\mathcal{S}_T &= \int\,d\tau\int\frac{d^3k}{(2\pi)^3}\sum_{i=1}^{2}\left[\frac{1}{2}|\partial_0 A_{\vec k}^{T_i}|^2-\frac{1}{2}(k^2+a^2m^2)|A_{\vec k}^{T_i}|^2\right]\nonumber\\
\mathcal{S}_L &= \int\,d\tau\int\frac{d^3k}{(2\pi)^3}\left[\frac{1}{2}\frac{a^2m^2}{k^2+a^2m^2}|\partial_0 A_{\vec k}^{L}|^2-\frac{1}{2}a^2m^2|A_{\vec k}^{L}|^2\right]\,,
\label{eq:transLongAction}
\end{align}
where $A_{\vec k}^{T_i}$ and $A_{\vec k}^L$ are the transverse and longitudinal modes respectively. We may then find the equations of motion for each 
\begin{align}
(\partial_0{}^2+k^2+a^2m^2)A_{\vec k}^T&=0\nonumber\\
\left(\partial_0{}^2+k^2+a^2m^2+\frac{1}{6}\frac{k^2a^2}{k^2+a^2m^2}R+3\frac{k^2a^4m^2}{(k^2+a^2m^2)^2}H^2\right)\chi_{\vec k}^L&=0\,
\label{eq:vectorEoM}
\end{align}
where we have rescaled the longitudinal component $A_{\vec k}^L$ such that it has a canonically normalized kinetic term,
\begin{equation}
A^L_{\vec k}=\sqrt{\frac{k^2+a^2m^2}{a^2m^2}}\chi_{\vec k}^L\,.
\label{eq:longitudinalRescaling}
\end{equation}
Note that the equation of motion for the transverse modes is exactly that of a massive, conformally coupled scalar, and while the equation of motion of the longitudinal mode is unwieldy, we can see that its early time limit gives
\begin{equation}
(\partial_0{}^2+k^2)\chi^L=0,
\label{eq:earlyLongMode}
\end{equation}
and we can apply the Bunch-Davies boundary conditions to obtain an asymptotic form of the mode function. Importantly, the leading asymptotic behavior is identical to the minimally coupled massless scalar. Each polarization has a polarization vector $\varepsilon_{m}^s$, where $s$ denotes the transverse ($T_1, T_2$) or longitudinal ($L$) component, which satisfy $\varepsilon_{i}^{s}\varepsilon_{i}^{s'}=\delta_{s,s'}$ and $\varepsilon_{i}^{L}=\hat{k}$.

For simplicity, we will consider the propagators of the transverse and longitudinal modes separately. The quantized transverse modes can be written
\begin{align}
A^{T,+}_{\vec k} &\rightarrow \frac{1}{\sqrt{2}(2\pi)^{3/2}}\left[e^{-\epsilon\tau}\tau^{1/2}H^{(2)}_\nu(k\tau)a^{T}_{\vec k}+e^{\epsilon\tau}\tau^{1/2}H_\nu^{(1)}(k\tau)a_{\vec k}^{T\dagger}\right]\nonumber\\
A^{T,-}_{\vec k} &\rightarrow \frac{1}{\sqrt{2}(2\pi)^{3/2}}\left[e^{\epsilon\tau}\tau^{1/2}H^{(2)}_\nu(k\tau)a^{T}_{\vec k}+e^{-\epsilon\tau}\tau^{1/2}H_\nu^{(1)}(k\tau)a_{\vec k}^{T\dagger}\right]
\end{align}
where $\nu^2=1/4-m^2/H^2$. From these fields, we can build the in-in propagators
\begin{align}
G^{(T)++}_{ij}(\tau,\tau') &= \Pi_{ij}\frac{(\tau\tau')^{1/2}}{2}\left[\Theta(\tau-\tau')e^{-\epsilon(\tau-\tau')}H^{(1)}_\nu(k\tau')H^{(2)}_\nu(k\tau) + \Theta(\tau'-\tau)e^{-\epsilon(\tau'-\tau)}H^{(1)}_\nu(k\tau)H^{(2)}_\nu(k\tau')\right]\nonumber\\
G^{(T)--}_{ij}(\tau,\tau') &= \Pi_{ij}\frac{(\tau\tau')^{1/2}}{2}\left[\Theta(\tau-\tau')e^{-\epsilon(\tau-\tau')}H^{(1)}_\nu(k\tau)H^{(2)}_\nu(k\tau') + \Theta(\tau'-\tau)e^{-\epsilon(\tau'-\tau)}H^{(1)}_\nu(k\tau')H^{(2)}_\nu(k\tau)\right]\nonumber\\
G^{(T)-+}_{ij}(\tau,\tau') &= \Pi_{ij}\frac{(\tau\tau')^{1/2}}{2}e^{\epsilon(\tau+\tau')}H^{(1)}_\nu(k\tau')H^{(2)}_\nu(k\tau)\nonumber\\
G^{(T)+-}_{ij}(\tau,\tau') &= \Pi_{ij}\frac{(\tau\tau')^{1/2}}{2}e^{\epsilon(\tau+\tau')}H^{(1)}_\nu(k\tau)H^{(2)}_\nu(k\tau')\,.
\label{eq:transverseProps}
\end{align}
where $\Pi_{ij}=\delta_{ij}-k_ik_j/k^2$. Tensor structure aside, these have the same form as the scalar propagators of Eq.~\eqref{eq:explicitScalarProps}.

The propagators of the longitudinal mode are similarly constructed, albeit with different mode functions, and has propagators which behave similarly to the transverse propagators in the early time limit, as shown by Eq.~\eqref{eq:earlyLongMode}. That the propagators have the same early time dependence on $\tau$, and thus $\epsilon$, is a direct consequence of the Bunch-Davies boundary conditions. The tensor structure for the longitudinal modes is also different, $k_ik_j/k^2$ in each propagator.

We will now consider the massless case. The gauge invariance of the action is retained even when we integrate out $A_0$ in Eq.~\eqref{eq:explicitVectorAction} with $m^2=0$. The field can again be split into two transverse and one longitudinal mode, but we see that the longitudinal action in Eq.~\eqref{eq:transLongAction} is proportional to $m^2$ and so is identically 0 in the massless case. This is a manifestation of gauge invariance: we are free to choose the longitudinal mode via a gauge transformation $A_i\rightarrow A_i+k_i\varphi$ for any scalar $\varphi$ without contributing to the action\footnote{One might prefer the reasoning that the longitudinal mode can no longer be considered physical because it is non-normalizable, as can be seen in Eq.~\eqref{eq:longitudinalRescaling} with $m^2=0$.}. For concreteness, we use a particularly useful gauge in FRW spaces, axial gauge, with $A^0=0$. This gives us the same equation of motion for the transverse components (Eq.~\eqref{eq:vectorEoM}) with $m=0$. The mode functions, and thus the propagators, simplify
\begin{align}
G^{++}_{ij}(\tau,\tau') &= \Pi_{ij}\frac{e^{-ik|\tau-\tau'|}}{2k}e^{-\epsilon|\tau-\tau'|}\nonumber\\
G^{--}_{ij}(\tau,\tau') &= \Pi_{ij}\frac{e^{ik|\tau-\tau'|}}{2k}e^{-\epsilon|\tau-\tau'|}\nonumber\\
G^{-+}_{ij}(\tau,\tau') &= \Pi_{ij}\frac{e^{-ik(\tau-\tau')}}{2k}e^{\epsilon(\tau+\tau')}\nonumber\\
G^{+-}_{ij}(\tau,\tau') &= \Pi_{ij}\frac{e^{ik(\tau-\tau')}}{2k}e^{\epsilon(\tau+\tau')}\,.
\label{eq:gaugeProps}
\end{align}
One may find the propagators for any other gauge by doing a gauge transformation, and reintroducing the $A_0$ via Eq.~\eqref{eq:A0Solution}. Because there is no constraint on the scalar function in the gauge transformation, we cannot guarantee that the propagators in other gauges will manifestly obey the criteria of our argument. However, because the physical degrees of freedom, the transverse modes, do have the proper early-time behavior, any perturbative computation of a physical observable with the perturbation series in Eq.~\eqref{eq:manifestUnitaryPT} will be equivalent to that computed by standard in-in perturbation theory.

In each of Eqs.~\eqref{eq:transverseProps}, \eqref{eq:gaugeProps} and the longitudinal propagators described above, the early-time behavior is identical to that of the scalar propagators in Eq.~\eqref{eq:explicitScalarProps} as expected, the only modifications being in the added tensor structure, which is explicitly time-independent. In any diagram, the tensor structures will be contracted with 3-momenta or the spatial metric, but this can only change the time-dependence of resulting integrals by powers of the scale factor. Since we only care about exponential $\epsilon$ dependence, such changes do not interfere with our argument.

\subsection{Spin-2}\label{subsec:graviton}

The inflationary period of our universe is well-approximated by a de Sitter geometry. To describe metric perturbations during inflation, the metric can be split into scalar, vector, and tensor components (though vectors are usually ignored) within the ADM formalism. In this formalism, there is a scalar degree of freedom left in the metric which will be combined with the inflaton field to create the observed gauge-invariant scalar fluctuations. There are two remaining degrees of freedom in the tensor modes, which then correspond to the graviton, and fluctuations of these degrees of freedom are gauge invariant.\footnote{Instead of using the ADM formalism, we could start with the standard graviton Lagrangian and choose the axial gauge for gravitons, $h_{0\mu}=0$, which would lead to the same results, analogously to the discussion of the massless vector in Sec.~\ref{subsec:vector}.} Mathematically, we can write the metric the following way~\cite{Maldacena03}
\begin{equation}
ds^2 = N^2\,dt^2-\gamma_{ij}(N^i\,dt+dx^i)(N^jdt+dx^j),
\end{equation}
where $N$ is the lapse function, $N^i$ is the shift vector, and $\gamma_{ij}$ is the spatial part of the metric, which still contains the scalar degree of freedom and the scale factor. We can write the spatial part of the metric as
\begin{equation}
\gamma_{ij} = a^2(t)(\zeta\delta_{ij}+h_{ij})
\end{equation}
where $t$ is cosmological time, and we can now see explicitly the scalar degree of freedom $\zeta$ and the gauge-invariant tensor fluctuation $h_{ij}$, which is a symmetric, transverse, and traceless $3\times3$ matrix. We will now consider the tensor fluctuations as a quantum field.

Tensor fluctuations of the metric have a quadratic action that looks like two massless minimally-coupled scalars, one for each helicity~\cite{BHT21}.
\begin{equation}
S_2=\frac{M_\text{pl}}{8}\int\,d^4x\,a^4(\tau)\left[\partial_0 h_{ij}\partial_0 h_{ij}-\partial_l h_{ij}\partial_l h_{ij}\right],
\label{eq:spin2Action}
\end{equation}
where repeated latin (spatial) indices are contracted by $\delta_{ij}$ and we are now using conformal time. The two polarization tensors $\varepsilon_{ij}^{+,\times}(\vec{k})$ have the properties $\varepsilon_{ij}^{s}(\vec{k})=\varepsilon_{ji}^{s}(\vec{k})$, $k_i\varepsilon_{ij}^{s}(\vec{k})=\varepsilon_{ii}^{s}(\vec{k})=0$ and $\varepsilon_{ij}^{s}(-\vec{k})\varepsilon_{ij}^{s'}(\vec{k})=4\delta_{s,s'}$. The mode functions are those of massless minimally-coupled scalars
\begin{equation}
h_{\vec k}(\tau)=\frac{H}{M_\text{pl}\sqrt{2k^3}}(1+ik\tau)e^{-ik\tau}\,.
\end{equation}
\begin{align}
h_{\vec k}^+ &\rightarrow \frac{H}{M_\text{pl}\sqrt{2k^3}}\left[e^{-\epsilon\tau}(1+ik\tau)e^{-ik\tau}a_{\vec k}+e^{\epsilon\tau}(1-ik\tau)e^{ik\tau}a_{\vec k}^{\dagger}\right]\nonumber\\
h_{\vec k}^+ &\rightarrow \frac{H}{M_\text{pl}\sqrt{2k^3}}\left[e^{\epsilon\tau}(1+ik\tau)e^{-ik\tau}a_{\vec k}+e^{-\epsilon\tau}(1-ik\tau)e^{ik\tau}a_{\vec k}^{\dagger}\right]
\end{align}
We also have propagators
\begin{align}
G^{++}_{ijmn}(\tau,\tau') &= \Pi_{ijmn}\frac{H^2}{M_\text{pl}^2}\frac{e^{-ik|\tau-\tau'|}}{2k^3}(1+ik|\tau-\tau'|+k^2\tau\tau')e^{-\epsilon|\tau-\tau'|}\nonumber\\
G^{--}_{ijmn}(\tau,\tau') &= \Pi_{ijmn}\frac{H^2}{M_\text{pl}^2}\frac{e^{ik|\tau-\tau'|}}{2k^3}(1-ik|\tau-\tau'|+k^2\tau\tau')e^{-\epsilon|\tau-\tau'|}\nonumber\\
G^{-+}_{ijmn}(\tau,\tau') &= \Pi_{ijmn}\frac{H^2}{M_\text{pl}^2}\frac{e^{-ik(\tau-\tau')}}{2k^3}(1-ik(\tau-\tau')+k^2\tau\tau')e^{\epsilon(\tau+\tau')}\nonumber\\
G^{+-}_{ijmn}(\tau,\tau') &= \Pi_{ijmn}\frac{H^2}{M_\text{pl}^2}\frac{e^{ik(\tau-\tau')}}{2k^3}(1+ik(\tau-\tau')+k^2\tau\tau')e^{\epsilon(\tau+\tau')}\,.
\label{eq:tensorProps}
\end{align}
where $\Pi^{ij}_{mn}=\Pi^{(i}_{(m}\Pi^{j)}_{n)}-\frac{1}{2}\Pi^{ij}\Pi_{mn}$, with $\Pi_{ij}$ defined in Eq.~\eqref{eq:transverseProps}. Like with the spin-1 case, the early time behavior is exactly that of a scalar, and the tensor structure cannot introduce non-trivial $\epsilon$-dependence when contracted with metrics and momenta.

\section{Discussion}\label{sec:conclusion}

It is relatively straightforward to adapt the arguments of this work to show that Weinberg's commutators are equivalent to usual in-in perturbation theory in more arbitrary inflationary spacetimes, including slow-roll spacetimes. In particular, any FLRW spacetime with a scale factor of the form $a(\tau)\sim\sum_{n=0}^{\infty}b_n \tau^{-n-s}$ for some $0<s\leq1$ at early times will follow the same logic as in this work. To see this, we note that the equations of motion in Eqs.~\eqref{eq:scalarEoM}, \eqref{eq:chiDiffEq}, and \eqref{eq:vectorEoM} are valid for any FLRW spacetime. Because the vector and scalar equations of motion involve products of scale factors with $H$ and $R$, we should make sure that these terms die off at early times. In conformal time, $H=\dot{a}/a^2$ and $R=6\ddot{a}/a^3$. At early times, the scale factor $a\sim \tau^{-s}$, so $H\sim\tau^{s-1}$ and $R\sim\tau^{2s-2}$. Since $s\leq1$, $H$ and $R$ behave at worst like constants, so when multiplied by further factors of the scale factor, these terms do indeed go to zero in the far past. Therefore, the equations of motion all reduce to the same form at lowest order in $1/\tau$ in the limit $\tau\rightarrow\infty$. This allows us to impose the Bunch-Davies boundary conditions on the asymptotic solutions to the equations of motion, similar to our treatment of the longitudinal mode of the massive vector in Sect.~\ref{subsec:vector}. The spinor and tensor structures for spins 1/2, 1, and 2 in FLRW are identical to those of the de Sitter case written in Sect.~\ref{subsec:fermion}, \ref{subsec:vector}, and \ref{subsec:graviton}.

It would be interesting to extend the ideas of this work to other classes of spacetimes, such as anti-de Sitter (AdS) or asymptotically-AdS spacetimes, where in-in perturbation theory can be used in explicit demonstrations of the AdS/CFT correspondence~\cite{Skenderis08}. There, the $i\epsilon$ prescription can be shown to be equivalent to an insertion of the vacuum wave function in the path integral. Additionally, a spin-independent proof of our result could be applied to the study of supersymmetry or higher-spin gravity theories. Higher-spin theories of gravity appear in realizations of the proposed dS/CFT correspondence~\cite{dSCFT}. Ultimately, one would want an extension of our proof which is spin-independent and works for any past-infinite causal spacetime.

%%%%%%%%%%%%%%%%%%%%%%%%%%%%%
\section*{Acknowledgements}
%%%%%%%%%%%%%%%%%%%%%%%%%%%%%

We thank M. Baumgart for many helpful comments on this work.

\bibliographystyle{JHEP}
\bibliography{draftv3}

\end{document}